\documentclass[12pt]{article}
\usepackage[pdftex]{color}
\usepackage{amsmath,amsthm,amssymb}
\usepackage{wasysym}
\usepackage{graphicx}
\usepackage{color}%
\usepackage[doublespacing]{setspace}
\usepackage{lipsum}
\usepackage[lmargin=2cm,rmargin=2cm,tmargin=2cm,bmargin=2cm]{geometry}
\usepackage{multirow}
\usepackage{hyperref}
\usepackage{mathtools}
\usepackage{breqn}
\usepackage{threeparttable}
\usepackage[utf8]{inputenc}
\usepackage{tocloft}
\usepackage{cite}
\usepackage{enumitem}
\usepackage{lineno}
\setlist[itemize]{leftmargin=*}

\begin{document}

\begin{center}
{\rm \bf \Large
Revisiting the diffusion equation derivation in Persson's theory of contact
}
\end{center}

\begin{center}
{\bf Yang Xu$^{\text{ab}}$\footnote{Corresponding author: yang.xu@hfut.edu.cn}, Siyuan Yang$^{\text{a}}$, Yunong Zhou$^{\text{c}}$, Liang Luo$^{\text{a}}$}
\end{center}

\begin{flushleft}
{
$^{\text{a}}$School of Mechanical Engineering, Hefei University of Technology, Hefei, 230009, China\\
$^{\text{b}}$Anhui Province Key Laboratory of Digital Design and Manufacturing, Hefei, 230009, China\\
$^{\text{c}}$Department of Civil Engineering, Yangzhou University, Yangzhou, 225127, China
}
\end{flushleft}

\begin{abstract}
In Persson's seminal work on tire-road interaction (Persson, J. Chem. Phys. {\bf 115}(8), 2001), he ingeniously derived a diffusion equation in Appendix B to characterize the evolution of contact pressure between a rigid rough indenter and an elastic half-space with spatial magnification, which constitutes the foundation of Persson's theory of contact. Persson's theory has been extensively validated and applied to investigate nearly all critical aspects of tribological problems. In contrast to the well-known Greenwood-Williamson (GW) model, Persson's theory receives relatively less attention within the tribology community. One contributing factor to this discrepancy is that the original derivation of the diffusion equation in Appendix B is not easily understandable to non-physicists. In this technical note, the authors provide supplementary information for each step of the derivation, aiming to clarify the conceptual foundation for researchers who encounter difficulties in understanding Persson's theory, thereby encouraging its broader application within and beyond tribology.
\end{abstract}

\section{Introduction}
Persson's theory of contact \cite{Persson01}, initially proposed in 2001, has since gained widespread recognition and application in the study of rough surface contact across various length scales, ranging from macroscopic to nanoscale. Utilizing Persson's theory of contact allows for an examination of the impacts of rough topographies at different scales on contact and frictional performance, thereby providing a robust theoretical foundation for addressing issues in diverse tribological applications \cite{persson2008theory, persson2025rubber, Persson22}. The core concept of this theory lies in characterizing the evolution of contact pressure distribution and gap distribution as roughness components with shorter wavelengths are progressively incorporated. In contrast to the GW model, which is extensively utilized in both academia and industry, Persson's theory receives relatively less attention from tribologists and solid mechanicians. A contributing factor to this limited attention is the overly simplified and inadequately explained derivation of the diffusion equation (Eqs. (B1-B5)), provided in Appendix B of Ref. \cite{Persson01}. Manners and Greenwood \cite{Manners06} undertook the first effort to ``decipher" Persson's theory, emphasizing the physical interpretation of the diffusion coefficient and deriving simplified closed-form solutions for the probability density function (PDF) of the contact pressure. Xu et al. \cite{xu2024persson} provided a tutorial elucidating the fundamentals of Persson's contact theory, which includes a novel and detailed derivation of the diffusion equation based on stochastic process theory. To assist researchers who still struggle to understand Persson's theory while reading his seminal 2001 work \cite{Persson01}, this technical note aims to bridge the gaps in the derivation of the diffusion equation outlined in Appendix B.

Before we revisit the rigorous derivation of the diffusion equation, we would like to emphasize that the diffusion equation can be directly deduced in a special case where the rough interface is completely flattened. According to the law of large numbers, the rough surface height follows a Gaussian distribution. Since the contact pressure is a convolution of the normal surface displacement (equivalently, the rough surface height), the contact pressure should also follow a Gaussian distribution, which is a special solution of a diffusion equation \cite{Manners06}. Xu et al. \cite{xu2014statistical, greenwood2015almost, Xu17} extended the multi-asperity contact models to the nearly complete contact ranges where the entire interface is almost completely flattened. The area-to-load relations predicted by both Xu et al's model and Persson's theory show good correlation. Ciavarella \cite{ciavarella2016rough} derived an asymptotic solution of real contact area based on Xu et al's model in the form of an error function, which is nearly identical to that of Persson's theory, except for a different constant multiplied to the error function. Even though the deduction of the diffusion equation from the complete contact assumption is straightforward, it is not a rigorous proof for the diffusion equation under partial contact.

\section{Problem statement}
\begin{figure}[h!]
  \centering
  \includegraphics[width=10cm]{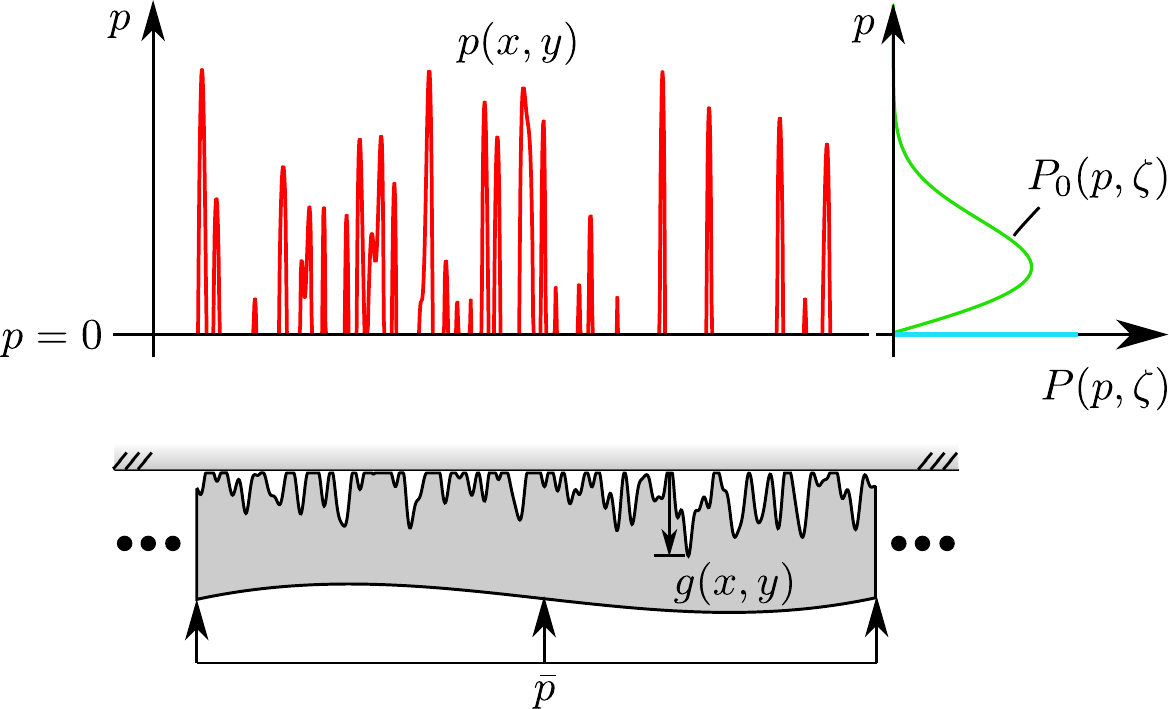}
  \caption{Schematic of the interfacial gap ($g(x,y)$), the elastic contact pressure ($p(x,y)$) and the corresponding PDF ($P(p, \zeta)$)}\label{fig:Fig_1}
\end{figure}

\begin{figure}[h!]
  \centering
  \includegraphics[width=14cm]{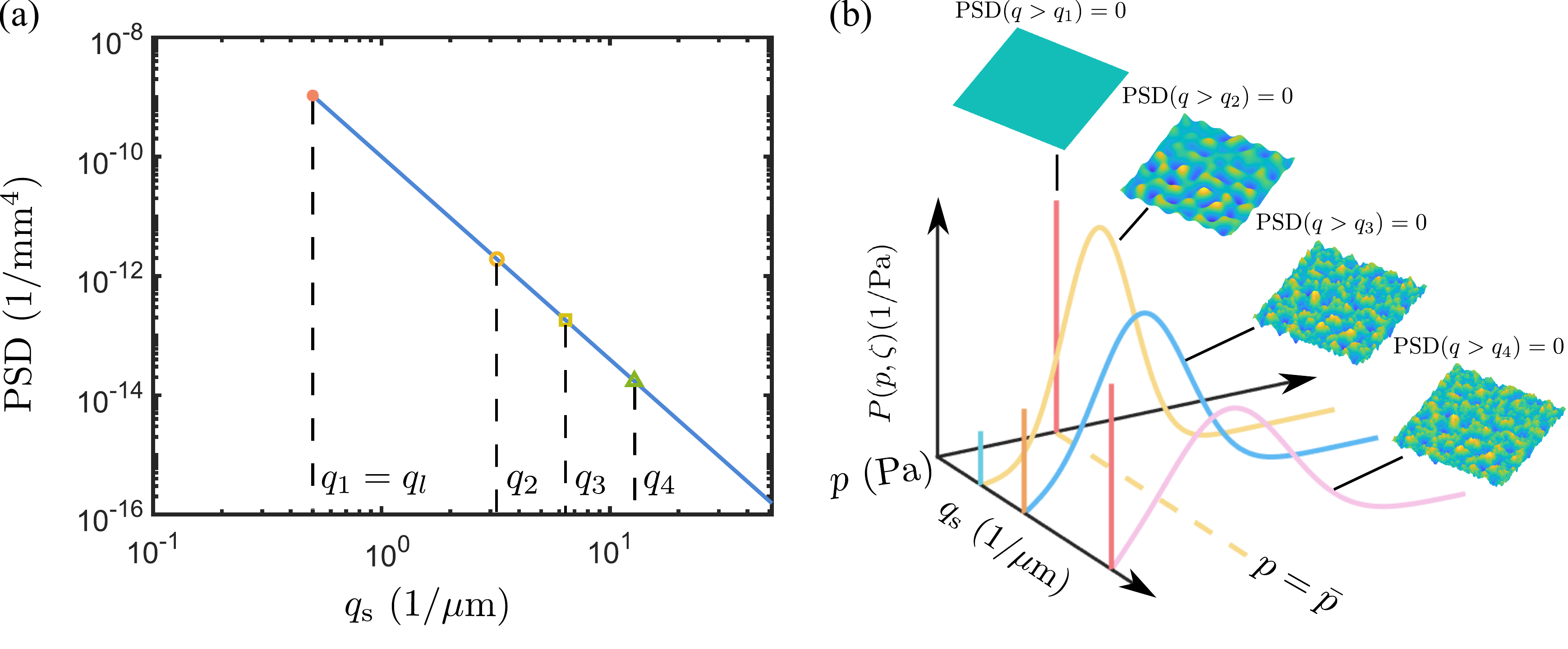}
  \caption{Schematics of (a) power spectral density and (b) the evolution of rough surface topography and PDF of contact pressure with magnification}\label{fig:Fig_2}
\end{figure}

The contact problem studied in this work is an indentation problem involving a rigid, nominally flat, rough surface in purely normal contact with an elastic half-space. The indenter is spatially fixed, while the half-space is subjected to a \textcolor{red}{purely normal, compressive, constant} traction ($\bar{p} > 0$) applied at a distance from the interface. The rough surface is random, bandwidth-limited, self-affine, isotropic, and nominally flat. The contact pressure distribution across the entire interface is denoted by $p(x, y) \geq 0$, where $(x, y) \in \mathbb{R}^2$. Moreover, shear stress, adhesion ($p < 0$), and the third body medium are neglected at the interface. A graphical illustration of the contact problem is provided in Fig. \ref{fig:Fig_1}.
The power spectral density (PSD) of the rough surface, denoted by $C(q_x, q_y)=C(q)$, follows a piecewise power law down to the atomic scale \cite{Gujrati18} (see Fig. \ref{fig:Fig_2}(a) for an example). 
Here, $q = \sqrt{q_x^2 + q_y^2}$ represents the wavenumber modulus of the wavenumber components in $x$ and $y$ directions (the reciprocal of the wavelength). 
The non-zero PSD is confined to a finite wavenumber range $q \in [q_{\text{l}}, q_{\text{s}}]$, where $q_{\text{l}}$ and $q_{\text{s}}$ correspond to the lower and upper cut-off wavelengths, respectively. 
The magnification $\zeta = q_{\text{s}}/q_{\text{l}}$ quantifies the multi-scale nature of the rough surface. As $\zeta$ increases from unity, the surface morphology transitions from a flat plane ($\zeta = 1$) to progressively rougher topography ($\zeta > 1$). 
This transition is accompanied by characteristic changes in the contact pressure distribution, which can be conveniently characterized by its PDF, $P(p, \zeta)$. 
It transforms from a Dirac delta function $\delta(p - \bar{p})$ at $\zeta = 1$ to a bell-shaped distribution with an increasing span along the pressure range. 
Notably, since the probability of $p > 0$ is not strictly unity, $P(p, \zeta)$ maintains a delta function-like peak at $p = 0$, see Fig. \ref{fig:Fig_2}(b) for an example. 

Based on the preceding description, $P(p, \zeta)$ comprises two distinct components: (1) the probability of zero pressure in the non-contact area and (2) the probability of compressive pressure in the contact area. This leads to the following piecewise representation:
\begin{equation}\label{eq:PDF_piecewise}
  P(p, \zeta) = \delta(p) \left[1 - A^*(\zeta) \right] + P_0(p, \zeta),
\end{equation}
where $p \in \mathbb{R}$, and the Dirac delta function $\delta(p)$ captures the concentration of probability at $p = 0$. The function $P_0(p, \zeta)$ represents the PDF of contact pressure within the real contact area, while $A^*$ denotes the magnification-dependent relative contact area, defined as the ratio of the real area of contact ($A_{\text{r}}$) to the nominal contact area ($A_{\text{n}}$):
\begin{equation}\label{eq:area}
A^*(\zeta) = \int_{0}^{\infty} P_0(p, \zeta) dp.
\end{equation}
The formulation strictly satisfies the probability conservation $\int_{-\infty}^{\infty} P(p, \zeta) dp = 1$.
%
\section{Diffusion equation for contact pressure}\label{sec:Diff_Persson}
The core concept of Persson's theory lies in the diffusion equation that describes the broadening of $P_0(p, \zeta)$ as $\zeta$ increases. Although Persson provided the derivation in Appendix B of Ref. \cite{Persson01} in only five steps (Eqs. (B1–B5)), the authors argue that a detailed, step-by-step explanation is necessary to ensure a complete understanding of the theory, particularly for non-physicists. To align with conventions in contact mechanics, the contact pressure symbol ($\sigma$) used in Persson's original derivation has been replaced by $p$. Notice that $P(\sigma, \zeta)$ used in Persson's theory corresponds to $P_0(p, \zeta)$ in the present work. In the rest of this section, we will first list the original equation provided by Persson in Appendix B of Ref. \cite{Persson01}, and the corresponding supplementary information will be given right after it.
\begin{equation}
    P_0(p, \zeta) = \langle \delta (p - p({\boldsymbol x}, \zeta)) \rangle, \tag{B1}
\end{equation}
{\bf Supplementary information} Due to the stochastic nature of the surface topography, the contact pressure at a given in-plane coordinate ${\boldsymbol x}=(x, y)$ and magnification $\zeta$ is a \emph{random} variable. In Persson's original derivation, the magnification symbol in $p({\boldsymbol x}, \zeta)$ was neglected for simplicity. In this study, $\zeta$ is retained to indicate that the contact pressure distribution is magnification-dependent.

Equation (B1) represents an ``unusual" method of defining probability density in the engineering community; nevertheless, it is frequently employed in statistical mechanics \cite{Risken89}. The derivation of Eq. (B1) as given in Chapter 2 of Risken's book \cite{Risken89} is briefly introduced below: Using the following definition of a Heaviside function:
\begin{equation}
  H(p - p({\boldsymbol x}, \zeta)) =
    \begin{cases}
      1   & ~ p({\boldsymbol x}, \zeta) \leq p,   \\
      0   & ~ p({\boldsymbol x}, \zeta) > p,
    \end{cases}
\end{equation}
the cumulative probability of the contact pressure with $0 < p({\boldsymbol x}, \zeta) \leq p$ is defined as
\begin{equation}\label{E:Probability_discretized}
    \Phi(p, \zeta) = \lim_{N\to \infty} \frac{1}{N} \left[ H(p - p({\boldsymbol x}_1, \zeta)) + H(p - p({\boldsymbol x}_2, \zeta)) + \ldots + H(p - p({\boldsymbol x}_N, \zeta))\right],
\end{equation}
based on the fundamental definition of probability. By examining the magnitude of the contact pressure $p({\boldsymbol x}, \zeta)$ at a significant number of locations ${\boldsymbol x}_i, i = 1, \dots, N$, the summation of $H(p - p({\boldsymbol x}_i, \zeta)), i = 1, \dots, N$ corresponds to the total number of points within the real contact area that satisfy $0 < p({\boldsymbol x}, \zeta) \leq p$. In accordance with the law of large numbers, as $N$ approaches infinity, the sample mean converges to the theoretical mean. Considering an ensemble of various rough surface realizations, we can utilize $\langle \ldots \rangle$ to substitute for $\displaystyle \lim_{N \to \infty} \frac{1}{N} \left[ \ldots \right]$, i.e.,
\begin{equation}
  \Phi(p, \zeta) = \langle H(p - p({\boldsymbol x}, \zeta)) \rangle.
\end{equation}
Using the definition of the probability density $\displaystyle P_0(p, \zeta) = \partial \Phi(p, \zeta)/\partial p$
and the identity $\displaystyle \delta(p) = dH(p)/dp$, we arrive at Eq. (B1).
\begin{align}
  P_0(p, \zeta + \delta \zeta) & = \langle \delta(p - p({\boldsymbol x}, \zeta) - \Delta p({\boldsymbol x}, \zeta) ) \rangle, \tag{B2a}\\
   & = \int_{-\infty}^{\infty} \langle \delta(\Delta p - \Delta p({\boldsymbol x}, \zeta)) \delta(p - p({\boldsymbol x}, \zeta)  - \Delta p) \rangle d\Delta p, \tag{B2b} \\
   & = \int_{-\infty}^{\infty} \langle \delta(\Delta p - \Delta p({\boldsymbol x}, \zeta)) \rangle P_0(p-\Delta p, \zeta) d\Delta p. \tag{B2}
\end{align}
{\bf Supplementary information} The $d\Delta p$ appears before the integrand in Eq. (B2) in Ref. \cite{Persson01}. This ``unusual" notation is not widely used in tribology and solid mechanics. It is a common practice adopted by physicists in statistical mechanics for managing lengthy integrands. However, this style will not be followed in this study, as the general audience is predominantly engineers. The upper and lower limits of the integral are not explicitly provided by Persson \cite{Persson01}. According to Manners and Greenwood \cite{Manners06}, as well as our derivation below, the upper and lower limits should be $-\infty$ and $\infty$, respectively. In Ref. \cite{Persson01}, the contact pressure distribution $p({\boldsymbol x}, \zeta)$ is simplified as $p$. However, it may easily be confused with the random variable of the same name. Therefore, the notation $p({\boldsymbol x}, \zeta)$ is retained for clarity in this study.

As $\zeta$ increases by $\Delta \zeta$, the contact pressure distribution evolves from $p({\boldsymbol x}, \zeta)$ to $p({\boldsymbol x}, \zeta) + \Delta p({\boldsymbol x}, \zeta)$. Following the definition of the probability density given in Eq. (B1), $P_0(p, \zeta + \Delta \zeta)$ is defined following Eq. (B2a). Equation (B2b) is derived using the convolution property of the Dirac delta function. This property states that the convolution of any function (say, $f(x)$) with the Dirac delta function is equivalent to the function itself, which can be expressed as follows:
\begin{equation}\label{eq:Dirac_convolution}
  (f*\delta)(x) \equiv \int_{-\infty}^{\infty} f(x')\delta(x-x') dx' = f(x).
\end{equation}
Let $x = p - p({\boldsymbol x}, \zeta) - \Delta p({\boldsymbol x}, \zeta)$, $x' = \Delta p - \Delta p({\boldsymbol x}, \zeta)$, $f(x') = \delta(x')$, and using the identity in Eq. \eqref{eq:Dirac_convolution},
\begin{equation}
\label{E:B2b_1}
  \delta(p - p({\boldsymbol x}, \zeta) - \Delta p({\boldsymbol x}, \zeta)) =
  \int_{-\infty}^{\infty} \delta(\Delta p - \Delta p({\boldsymbol x}, \zeta))
  \delta(p - p({\boldsymbol x}, \zeta) - \Delta p)~d\Delta p.
\end{equation}
Applying the ensemble average to both sides of Eq. \eqref{E:B2b_1}, we obtain Eq. (B2b). 

The ensemble term, $\langle \delta(\Delta p - \Delta p({\boldsymbol x}, \zeta)) \delta(p - \Delta p - p({\boldsymbol x}, \zeta)) \rangle$, on the right-hand side (R.H.S) of Eq. (B2b) is the same as the \emph{joint} PDF $P_0(\Delta p, p - \Delta p, \zeta)$, provided that $\Delta p$ and $p$ are two independent random variables. This can be proven as follows: Inspired by Eq. (B1), the joint cumulative probability $\Phi(\Delta p, p - \Delta p, \zeta)$ that satisfies $\Delta p({\boldsymbol x}, \zeta) \leq \Delta p, p({\boldsymbol x}, \zeta) \leq p - \Delta p$ can be written as follows:
\begin{align}
  \Phi(\Delta p, p - \Delta p, \zeta) = \lim_{N\to \infty} \frac{1}{N} [ &H(\Delta p - \Delta p({\boldsymbol x}_1, \zeta)) H(p - \Delta p - p({\boldsymbol x}_1, \zeta)) + \ldots + \notag \\
& H(\Delta p - \Delta p({\boldsymbol x}_N, \zeta)) H(p - \Delta p - p({\boldsymbol x}_N, \zeta)) ].
\end{align}
Replacing $\displaystyle{\lim_{N \to \infty} \frac{1}{N}}[\cdots]$ with the ensemble average,
\begin{equation}
  \Phi(\Delta p, p - \Delta p, \zeta) = \langle H(\Delta p - \Delta p({\boldsymbol x}, \zeta)) H(p - \Delta p - p({\boldsymbol x}, \zeta)) \rangle.
\end{equation}
According to the definition of the joint PDF $\displaystyle P_0(x, y) = \partial^2 \Phi(x, y)/\partial x \partial y$, 
\begin{equation}\label{E:B2c_1}
  P_0(\Delta p, p - \Delta p, \zeta) = \langle \delta(\Delta p - \Delta p({\boldsymbol x}, \zeta)) \delta(p - \Delta p - p({\boldsymbol x}, \zeta)) \rangle.
\end{equation}
When $\Delta p$ and $p$ are two independent random variables,
\begin{equation}
\langle \delta(\Delta p - \Delta p({\boldsymbol x}, \zeta)) \delta(p - \Delta p - p({\boldsymbol x}, \zeta)) \rangle = \langle \delta(\Delta p - \Delta p({\boldsymbol x}, \zeta)) \rangle  \langle \delta(p - \Delta p - p({\boldsymbol x}, \zeta)) \rangle. \label{E:B2c_2}
\end{equation}
Substituting Eq. \eqref{E:B2c_2} into Eq. (B2b) and using $\langle \delta(p - \Delta p - p({\boldsymbol x}, \zeta)) \rangle = P_0(p - \Delta p, \zeta)$, Eq. (B2) is derived.

Replacing $\langle \delta(\Delta p - \Delta p({\boldsymbol x}, \zeta)) \rangle$ in Eq. (B2) by $P_0(\Delta p, \zeta)$ allows us to further simplify Eq. (B2) to
\begin{equation}\label{E:B2c_3}
  P_0(p, \zeta + \Delta \zeta) = \int_{-\infty}^{\infty} P_0(\Delta p, \zeta) P_0(p - \Delta p, \zeta) d\Delta p,
\end{equation}
where $P_0(p, \zeta + \Delta \zeta)$ and $P_0(p - \Delta p, \zeta)$ are PDFs of the contact pressure at magnification levels $\zeta + \Delta \zeta$ and $\zeta$, respectively. $P_0(\Delta p, \zeta)$ is known as the \emph{transition} probability, representing the conditional probability of $p({\boldsymbol x}, \zeta + \Delta \zeta) = p$ given $p({\boldsymbol x}, \zeta) = p - \Delta p$. Equation \eqref{E:B2c_3} is recognized as the \emph{Chapman-Kolmogorov} equation. The stochastic process describing contact pressure $p({\boldsymbol x}, \zeta)$, defined by Eq. \eqref{E:B2c_3}, is called a \emph{Markov} process. A detailed derivation of the diffusion equation from the perspective of stochastic process theory has been given in Ref. \cite{xu2024persson}.

\begin{equation}
  \langle \delta(\Delta p - \Delta p({\boldsymbol x}, \zeta)) \rangle = \frac{1}{2\pi} \int_{-\infty}^{\infty} \langle e^{iw[\Delta p - \Delta p({\boldsymbol x}, \zeta)]} \rangle dw \tag{B3},
\end{equation}
where the integral limits were not explicitly given in Eq. (B3) of Ref. \cite{Persson01}. 

{\bf Supplementary information} The derivation of Eq. (B3) relies on an important concept: \emph{characteristic} function\cite{Risken89}. $P_0(\Delta p, \zeta)$ and its characteristic function form a Fourier transform pair:
\begin{equation}\label{E:B3_1}
  C(w, \zeta) = \int_{-\infty}^{\infty} e^{-i w \Delta p} P_0(\Delta p, \zeta) d\Delta p,
\end{equation}
\begin{equation}\label{E:B3_2}
  P_0(\Delta p, \zeta) = \frac{1}{2 \pi} \int_{-\infty}^{\infty} C(w, \zeta) e^{i w \Delta p} dw.
\end{equation}
Substituting $P_0(\Delta p, \zeta) = \langle \delta(\Delta p - \Delta p({\boldsymbol x}, \zeta)) \rangle$ into Eq. \eqref{E:B3_1},
\begin{equation}\label{E:B3_3}
  C(w, \zeta) = \langle \int_{-\infty}^{\infty} e^{-i w \Delta p} \delta(\Delta p - \Delta p({\boldsymbol x}, \zeta)) d \Delta p \rangle = \langle e^{-i w \Delta p({\boldsymbol x}, \zeta)} \rangle.
\end{equation}
Substituting Eq. \eqref{E:B3_3} into Eq. \eqref{E:B3_2}, we obtain Eq. (B3).

\begin{equation}
\langle \delta(\Delta p - \Delta p({\boldsymbol x}, \zeta)) \rangle = \frac{1}{2\pi} \int_{-\infty}^{\infty} e^{iw \Delta p} \left[1 - \frac{1}{2} w^2 \langle \Delta p^2({\boldsymbol x}, \zeta) \rangle \right] dw \tag{B4}.
\end{equation}

{\bf Supplementary information} The derivation of Eq. (B4) involves the Taylor series expansion of $e^{i w [\Delta p - \Delta p({\boldsymbol x}, \zeta)]}$ on the R.H.S of Eq. (B3). By expanding the exponential term, $e^{-i w \Delta p({\boldsymbol x}, \zeta)}$, about $\Delta p({\boldsymbol x}, \zeta) = 0$ up to the second order, we can obtain
\begin{align}\label{E:B4_1}
  e^{i w (\Delta p - \Delta p({\boldsymbol x}, \zeta))} & = e^{i w \Delta p} \left[ 1 - \Delta p({\boldsymbol x}, \zeta) i w  + \frac{1}{2!} (\Delta p({\boldsymbol x}, \zeta) i w )^2 - ...  \right], \notag \\
& \approx e^{i w \Delta p} \left[ 1 - \Delta p({\boldsymbol x}, \zeta) i w - \frac{1}{2} \Delta p^2({\boldsymbol x}, \zeta) w^2 \right].
\end{align}
Substituting Eq. \eqref{E:B4_1} into R.H.S of Eq. (B3), we obtain
\begin{equation}\label{E:B4_2}
  \text{R.H.S of Eq. (B3)} = \frac{1}{2\pi} \left[ \int_{-\infty}^{\infty} e^{iw\Delta p} dw -
  i \int_{-\infty}^{\infty} \langle \Delta p({\boldsymbol x}, \zeta) \rangle e^{i w \Delta p} w dw -
  \frac{1}{2} \int_{-\infty}^{\infty} \langle \Delta p^2({\boldsymbol x}, \zeta) \rangle e^{i w \Delta p} w^2 dw \right],
\end{equation}
where $\langle \Delta p({\boldsymbol x}, \zeta) \rangle$ and $\langle \Delta p^2({\boldsymbol x}, \zeta) \rangle$ represent the mean and variance of the incremental contact pressure $\Delta p({\boldsymbol x}, \zeta)$, respectively. The constant external load ($\bar{p}$) implies $\langle \Delta p({\boldsymbol x}, \zeta) \rangle = 0$. Thus, the second integral on the R.H.S of Eq. \eqref{E:B4_2} vanishes. Substituting Eq. \eqref{E:B4_2} with $\langle \Delta p({\boldsymbol x}, \zeta) \rangle = 0$ into Eq. (B3), we can obtain Eq. (B4). Now, the explicit form of the transition probability, $\langle \delta(\Delta p - \Delta p({\boldsymbol x}, \zeta)) \rangle$, is derived. 
\begin{equation}
  \frac{\partial }{\partial \zeta}P_0(p, \zeta) = \frac{1}{2} \lim_{\Delta \zeta \to 0} \frac{\langle \Delta p^2 \rangle}{ \Delta \zeta} \frac{\partial^2}{\partial p^2} P_0(p, \zeta), \tag{B5}
\end{equation}
where $\langle \Delta p^2 \rangle = \langle \Delta p^2({\boldsymbol x}, \zeta) \rangle$ for short.

{\bf Supplementary information} Substituting Eq. (B4) into Eq. (B2), we obtain
\begin{equation}\label{E:B5_1}
  P_0(p, \zeta + \Delta \zeta) = \int_{-\infty}^{\infty} \left[ \frac{1}{2 \pi} \int_{-\infty}^{\infty} e^{iw \Delta p} dw - \frac{1}{2}\langle \Delta p^2 \rangle \frac{1}{2 \pi} \int_{-\infty}^{\infty} w^2 e^{i w \Delta p} dw \right] P_0(p - \Delta p, \zeta) d\Delta p.
\end{equation}
The square bracket can be further simplified using Eq. (B3). Let $\Delta p({\boldsymbol x}, \zeta) = 0$ in Eq. (B3), then we obtain an identity:
\begin{equation}\label{E:B5_2}
\frac{1}{2\pi} \int_{-\infty}^{\infty} e^{iw\Delta p} dw = \delta(\Delta p).
\end{equation}
Applying $\displaystyle{\frac{\partial^2}{\partial (\Delta p)^2}}$ to both sides of Eq. (B3) and letting $\Delta p({\boldsymbol x}, \zeta) = 0$ yields another identity:
\begin{equation}
-\frac{1}{2 \pi} \int_{-\infty}^{\infty} w^2 e^{i w \Delta p} dw = \frac{d^2 \delta }{d (\Delta p)^2}. \label{E:B5_3}
\end{equation}
Substituting Eqs. \eqref{E:B5_2} and \eqref{E:B5_3} into Eq. \eqref{E:B5_1}, we obtain:
\begin{equation}\label{E:B5_4}
P_0(p, \zeta + \Delta \zeta) = P_0(p, \zeta) + \frac{1}{2} \langle \Delta p^2 \rangle \int_{-\infty}^{\infty} P_0(p - \Delta p, \zeta) \frac{d^2 \delta}{d (\Delta p)^2} d\Delta p.
\end{equation}
The integral in the equation above can be further simplified using integration by parts:
\begin{align}
   &\int_{-\infty}^{\infty} P_0(p - \Delta p, \zeta) \frac{d^2 \delta}{d (\Delta p)^2} d\Delta p \notag \\
   &= \left.P_0(p - \Delta p, \zeta) \frac{d \delta }{d \Delta p} \right\vert_{-\infty}^{\infty} - \left.\frac{\partial }{\partial \Delta p} P_0(p - \Delta p, \zeta) \delta(\Delta p) \right\vert_{-\infty}^{\infty} + \frac{\partial^2}{\partial p^2} P_0(p, \zeta) \label{E:B5_5}.
\end{align}
Since the Dirac delta function and its derivatives vanish at infinity in the above equation, Eq. \eqref{E:B5_4} becomes
\begin{equation}\label{E:B5_6}
P_0(p, \zeta + \Delta \zeta) = P_0(p, \zeta) + \frac{1}{2} \langle \Delta p^2 \rangle \frac{\partial^2}{\partial p^2} P_0(p, \zeta).
\end{equation}
Finally, Eq. (B5) is derived from Eq. \eqref{E:B5_6} with $\displaystyle\frac{\partial }{\partial \zeta} P_0(p, \zeta) = \lim_{\Delta \zeta \to 0} \frac{P_0(p, \zeta + \Delta \zeta) - P_0(p, \zeta)}{\Delta \zeta}$.

\section{Discussion}

A similar derivation can be applied to derive the partial differential equation for the interfacial gap distribution
$P_0(g, \zeta)$, which will be shown to be identical to the convection-diffusion equation derived by Xu et al. \cite{xu2024stochastic}. To ensure consistency with the derivation provided in Ref.  
\cite{Persson01}, the derivation is divided into five steps. The PDF of the interfacial gap, $P_0(g, \zeta)$, can be formulated using a Dirac delta function as follows: 
\begin{equation}
  P_0(g, \zeta) = \langle \delta(g - g({\boldsymbol x}, \zeta)) \rangle,
\end{equation}
where $g({\boldsymbol x}, \zeta)$ denotes the gap distribution within the non-contact area ($p = 0$) at the specified magnification $\zeta$. As $\zeta$ increases by $\Delta\zeta$, the gap distribution is expected to change to $g({\boldsymbol x}, \zeta) + \Delta g({\boldsymbol x}, \zeta)$. Thus, the PDF of $g({\boldsymbol x}, \zeta) + \Delta g({\boldsymbol x}, \zeta)$ is
\begin{equation}
\begin{aligned}
P_0(g, \zeta + \Delta\zeta) &=
  \langle \delta(g - g({\boldsymbol x}, \zeta) - \Delta g({\boldsymbol x}, \zeta))  \rangle, \\
  &=
  \int_{-\infty}^{\infty}
  \langle
  \delta(\Delta g - \Delta g({\boldsymbol x}, \zeta))
  \delta(g - \Delta g - g({\boldsymbol x}, \zeta) )
  \rangle d\Delta g, \\
  &=
  \int_{-\infty}^{\infty}
  \langle
  \delta(\Delta g - \Delta g({\boldsymbol x}, \zeta))
  \rangle
  P_0(g - \Delta g, \zeta) d\Delta g. \label{eq:C2}
\end{aligned}
\end{equation}
According to the definition of the Dirac delta function, we obtain
\begin{equation}
  \langle \delta(\Delta g - \Delta g({\boldsymbol x}, \zeta)) \rangle =
  \frac{1}{2\pi} \int_{-\infty}^{\infty}
  \langle
  e^{iw(\Delta g - \Delta g({\boldsymbol x}, \zeta))}
  \rangle dw.
\end{equation}
Expanding the exponential integrand up to the second order, 
\begin{equation}\label{eq:C4}
  \langle \delta(\Delta g - \Delta g({\boldsymbol x}, \zeta)) \rangle =
  \frac{1}{2\pi} \int_{-\infty}^{\infty}
  e^{iw\Delta g} \left[
    1 - iw\langle \Delta g({\boldsymbol x}, \zeta) \rangle - \frac{1}{2} w^2 \langle \Delta g^2({\boldsymbol x}, \zeta) \rangle
  \right] dw.
\end{equation}
Substituting Eq. \eqref{eq:C4} into Eq. \eqref{eq:C2} and letting $\Delta \zeta \to 0$,
\begin{equation}
\frac{\partial}{\partial \zeta}P_0(g, \zeta) = -\lim_{\Delta \zeta \to 0}
\frac{\langle \Delta g \rangle}{\Delta\zeta}\frac{\partial}{\partial g} P_0(g, \zeta) + \frac{1}{2} \lim_{\Delta \zeta \to 0} \frac{\langle \Delta g^2 \rangle}{\Delta\zeta} \frac{\partial^2}{\partial g^2}P_0(g, \zeta),
\end{equation}
where $\langle \Delta g \rangle$ and $\langle \Delta g^2 \rangle$ are abbreviations for $\langle \Delta g(\boldsymbol{x}, \zeta) \rangle$ and $\langle \Delta g^2(\boldsymbol{x}, \zeta)\rangle$. The two coefficients in front of the first and second partial derivative terms represent the drift and diffusion coefficients, respectively. 

The diffusion equation of the contact pressure, along with the boundary conditions $P_0(p = 0, \zeta) = P_0(p \to \infty, \zeta) = 0$ and the initial condition $P_0(p, \zeta = 1) = \delta(p - \bar{p})$, leads to the well-known mirrored Gaussian solution \cite{xu2024persson}:
\begin{equation}
P_0(p, \zeta) = \frac{1}{\sqrt{2 \pi V(\zeta)}} \left( \exp\left[ -\frac{(p - \bar{p})^2}{2 V(\zeta)} \right] - \exp\left[ -\frac{(p + \bar{p})^2}{2 V(\zeta)} \right] \right),
\end{equation}
where $V(\zeta) = \int_1^{\zeta} \langle \Delta p^2 \rangle(\zeta') d\zeta'$ is the variance of contact pressure. The extensively used error functional form of relative contact area can be immediately deduced from Eq. \eqref{eq:area} as follows \cite{xu2024persson}:
\begin{equation}
A^* = \text{erf}\left(\bar{p}/\sqrt{2 V(\zeta)}\right).
\end{equation}
A closed-form expression for the elastic strain energy $U_{\text{el}}$ stored in the deformed rough interface can be directly deduced from the above solutions \cite{Persson08}. Using the identity $\bar{p} = -\frac{1}{A_n}\partial U_{\text{el}}/\partial \bar{g}$, where $\bar{g}$ represents the average interfacial gap, a semi-logarithmic form of the $\bar{g}(\bar{p})$ relation at low pressure ranges can be derived \cite{Persson07}. Starting from $P_0(p, \zeta)$, the elastic strain energy \cite{Persson08}, the average interfacial gap \cite{Persson07}, elastoplastic contact \cite{lambert2025competition}, viscoelastic contact \cite{Papangelo21}, adhesion \cite{Persson02adhesion}, and friction \cite{Persson01} can be developed. The applications of Persson's theory in various fields of study (e.g., electrical contact \cite{Persson22}, seals \cite{bottiglione2009leakage}, and wear \cite{persson2025rubber}) become possible.

The diffusion equation can still be applicable in the case of adhesion. Persson has introduced an absorbing boundary condition $P_0(\sigma_{\text{a}}(\zeta))$ to allow for tensile traction. The critical value of $\sigma_{\text{a}}(\zeta)$ can be derived using a fracture mechanics approach based on the penny-shaped crack problem. Unlike the local energy balance approach proposed by Persson, the thermodynamic approach originally proposed by Johnson et al. \cite{johnson1971surface} (see also the work of Ciavarella et al. \cite{ciavarella2018approximate}) can be utilized for solving $\sigma_{\text{a}}(\zeta)$. 
By formulating the elastic strain energy, we can mimic the Griffith criterion using the identity $\partial U_{\text{el}}/\partial A_{\text{r}} = w$, where $w$ denotes the work of adhesion. The aforementioned adhesion models assume that surface interaction cannot occur outside the contact area. 
A more elegant variant of Persson's theory has been introduced by Joe et al. \cite{Joe17} based on stochastic process theory. An arbitrary inter-molecular interaction can be implemented in Persson's theory so that the solution of pressure PDF can be derived, ranging from the Johnson-Kendall-Roberts (JKR) limit to the Derjaguin-Muller-Toporov (DMT) limit.

\section{Conclusion}
In this technical note, we provide detailed supplementary information for Eqs. (B1-B5) in Appendix B of Ref. \cite{Persson01}. Two important prerequisites are introduced: (1) the probability expressed in terms of the Dirac delta function and (2) the characteristic function. A similar approach has been adapted to derive the convection-diffusion equation for the PDF of the interfacial gap. This technical note serves as complementary material for understanding Persson's original work \cite{Persson01} and paves the way for advanced applications of Persson's theory of contact in various fields. 

\section*{Acknowledgement}
This work was supported by the National Natural Science Foundation of China (No. 52105179, No. 12402116), the Fundamental Research Funds for the Central Universities (No. PA2024GDSK0044, No. JZ2025HGTG0298).

\begin{spacing}{1} 
	\bibliographystyle{asmejour}
	\bibliography{ref}
\end{spacing}

\end{document}